\documentclass[11pt,a4paper]{article} 
\usepackage[dvips]{graphicx}
\usepackage{amsfonts,amsmath,amssymb}
\usepackage{hyperref}

\begin{document}
\title{Exact Solutions of a Fermion-Soliton System in Two Dimensions}

\author{
L. Shahkarami\footnote{Electronic address: l$\_$shahkarami@sbu.ac.ir}~~and S.S. Gousheh\\
 \small Department of Physics, Shahid Beheshti University G.C., Evin, Tehran
19839, Iran}
\maketitle
\begin{abstract}
We investigate a coupled system of a Dirac particle and a
pseudoscalar field in the form of a soliton in ($1+1$) dimensions
and find some of its exact solutions numerically.
 We solve the coupled set of equations self-consistently and non-perturbatively by the use of a numerical method and obtain
the bound states of the fermion and the shape of the soliton. That
is the shape of the static soliton in this problem is not prescribed
and is determined by the equations themselves. This work goes beyond
the perturbation theory in which the back reaction of the fermion on
soliton is its first order correction. We compare our results to those
of an exactly solvable model in which the soliton is prescribed. We show that, as
expected, the total energy of our system is lower than the prescribed one. We also compute non-perturbatively
the vacuum polarization of the fermion induced by the presence of the soliton and display the results. Moreover, we compute the soliton
mass as a function of the boson and fermion masses and find that the
results are consistent with Skyrme's phenomenological conjecture.
Finally, we show that for fixed values of the parameters, the shape
of the soliton obtained from our exact solutions depends slightly on
the fermionic state to which it is coupled. However, the exact shape of the soliton is always very close to the
isolated kink.
\end{abstract}

\section{Introduction}
Systems consisting of coupled fermionic and bosonic fields,
specially when the latter is in the form of a soliton or a solitary
wave, have played an important role in many branches of physics.
Since solitons are non-dispersive localized packets of energy, they
present the possibility of describing extended objects, such as
hadrons as stable states within quantum field theory. For example,
starting with the work of Skyrme \cite{skyrme,skyrme2,skyrme3,skyrme4,skyrme5} in
1958, much work has been done for describing hadrons and their
interactions in the non-perturbative QCD regime using
phenomenological non-linear field theories. The simplest examples of
such models are the variety of bag models
\cite{mit,bagmodel1,bagmodel2,slac,bagmodel3,bagmodel4,bagmodel5,bagmodel6}
in some of which the effect of gluons is replaced by the interaction
of scalar fields with the quarks \cite{scalar,scalar2}.
For example static bag models such as MIT and SLAC models have had
notable success in describing hadronic structure. In the papers
on the SLAC bag model, first proposed by \cite{slac}, several hadronic properties
including their quantum numbers, form factors, as well as the
bound-state energies are calculated.

A coupled fermion-soliton system also appears in a set of works that
examine the effects of quantum corrections on solitons and the
formation of a quantum soliton which is not present at the classical
level \cite{heavy,heavyb,heavy2,heavy3,heavy4}. 
In these works, the authors
consider a theory of a scalar field coupled to a heavy fermion, i.e.
a strongly coupled fermion-scalar system. In these models fermions
acquire mass through their couplings to a scalar field with
non-vanishing vacuum expectation value. According to this assertion,
a strong fermion-scalar coupling is analogous to the existence of
heavy fermions. Considering the zero and first order quantum
corrections, they showed that heavy fermions can stabilize the
soliton.

Fermion-soliton systems also appear in the context of the so-called
braneworld models. A major issue in these models is the localization
problem \cite{localize,localize2,localize3} of the Standard Model
fields on the brane. Localizing a fermion on the brane was first
described by the original work of Rubakov and Shaposhnikov
\cite{rubakov}, in which a 5-dimensional model, i.e. one with
codimension-1, was employed and the brane Lagrangian was the
$\phi^4$ (or sine-Gordon) system in one dimension, supporting the
kink soliton. Extension to higher dimensions for this mechanism has
been performed by other authors
\cite{highercodim,highercodim2,highercodim3,highercodim4}. In the
following years, there has been some considerable activities in this
context \cite{localization,localization2,localization3}.

A more curious phenomenon when a fermion interacts with a soliton
is the assignment of half-integer fermion number to the solitonic
states. This was first pointed out by Jackiw and Rebbi
\cite{jakiew}. They considered the structure of a coupled
fermion-soliton system which possesses CP symmetry and showed that
the existence of a zero-energy fermion mode, implies that the
soliton is a degenerate doublet carrying fermion number $\pm
\frac{1}{2}$. The consequences of this surprising result have been
studied extensively for many different physical phenomena in the
literature, such as in the context of condensed-matter systems
\cite{condmatt,condmatt2}, polyacetylene being the standard example
\cite{polyace,polyace2}, and in high energy physics. Goldstone and
Wilczek \cite{goldstone} introduced a powerful method, called the
adiabatic method, for calculating the vacuum polarization of the
Fermi field induced by the soliton. In this method, the
topologically non-trivial configuration of the background scalar
field which is coupled to the fermions, is imagined to evolve
continuously and slowly from a topologically trivial configuration.
With the aid of their adiabatic method, they concluded that when the
system does not possess CP symmetry, the fermion number of the
soliton can be any real value, not just $\pm \frac{1}{2}$. Later on
the computational technique of the adiabatic method was modified by
MacKenzie and Wilczek, such that the requirement of adiabaticity was
lifted. In their method one computes the energy spectrum of the
fermion in the presence of a prescribed soliton and deduces the
induced vacuum polarization. They used it to illustrate the vacuum
polarization of fermion fields by solitons \cite{mackenzie1} and
then applied it for the case of infinitely sharp soliton as an
example \cite{mackenzie2}. Using this method, the vacuum
polarization by solitons for an exactly solvable model was computed
by Gousheh and Mobilia \cite{dr}.
 The model was general enough to include both the adiabatic and non-adiabatic cases,
  and they showed that even in the non-adiabatic cases solitons
can polarize the vacuum. Only the infinitely sharp solitons can
never polarize the vacuum.

The simplest example of a static topological solitary wave in
($1+1$) dimensions is the kink of the $\phi^4$ model which has many
applications in quantum field theory \cite{Manton}, condensed-matter
physics \cite{Bishop}, and cosmology \cite{Vilenkin}. It is worth
mentioning that since in the literature the distinction between
solitary waves and solitons is not highlighted and they are used
interchangeably, we will do the same in this paper. The model we
investigate is a ($1+1$)-dimensional theory in which a pseudoscalar field within the
non-linear sigma model is coupled to a fermion. The value of the pseudoscalar field at $x=\pm \infty$,
 denoted by $\pm\theta_0$, is a free parameter in the model. 
However, we are eventually interested
 in a model in which the pseudoscalar field is a solitary wave,
  i.e. we let $\theta_0$ evolve from $0$ to $\pi$, with \textit{apriori} undetermined shape. Although, many
aspects of this model, including the quantum corrections, have been
investigated in the literature, many important issues at the zero
order are yet to be investigated. The purpose of this paper is to
precisely do that. We solve the classical field equations of this
theory exactly, within numerical approximation, and extract many
important physical implications. The numerical method that we use is
the relaxation method. We also compare our results with two
interesting models. One of them is the exactly solvable model presented
by Gousheh and Mobilia \cite{dr}, in which a prescribed piecewise
linear pseudoscalar field with arbitrary boundary conditions has
been used and the complete fermionic eigenfunctions and eigenvalues have been
obtained. The shape of the pseudoscalar field is chosen so as to
qualitatively approximate the isolated kink. The second one is similar to the first,
except for the replacement of the piecewise linear field by the
isolated kink solution as a prescribed background field. The latter
is not exactly solvable and we solve it by using the same numerical
method as our model and, as expected, its results are very similar
to our model. We find that the exact (relaxed) shape of the soliton
in our model is very close to that of the isolated kink. However,
the spectrum of the fermion is completely altered due to the
presence of the interaction with the background field. We also find that
the total energy in the relaxed model is lower than the exactly solvable model with the prescribed soliton. 
Our solutions indicate that when the mass of the fermion $M$ and soliton
are equal, the mass of the elementary boson is $m\approx0.116 M$,
thus giving a strong indication for the validity of the Skyrme's
phenomenological model. We also find the surprising result that the
true (relaxed) shape of the soliton depends slightly on the
fermionic level considered. Moreover, comparison of the exact
results of our system with those of the system with the prescribed
kink, shows that the latter is a very good approximation for the
former. In section 2 we introduce the coupled fermion-soliton model, find the dynamical
equations, discuss the symmetries of the system and how they enter the boundary conditions.
We also discuss very briefly the numerical method used. 
In section 3 we discuss the results,
including the pattern of the fermion bound state energies and wave functions, the shape of the
solitons, and the total energies for the three models.
\section{The model of a coupled fermion-soliton system}
Consider the following Lagrangian describing a fermion interacting
non-linearly with a pseudoscalar field $\phi\left(x,t\right)$ in
($1+1$) dimensions:
\begin{eqnarray}\label{lagrangian den.}\vspace{.2cm}
  {\cal L}&=& {\bar{\psi}}\left(x,t\right)\left[i \gamma^\mu \partial_\mu
 - M \textit{e}^{i\phi\left(x,t\right)\gamma^5}\right]\psi \left(x,t\right)
\nonumber\\&-&\frac{{1}}{{4}}\lambda \left[\phi^2\left(x,t\right) -
\frac{m^2}{\lambda}\right]^2+\frac{{1}}{{2}}
\partial_{\mu}\phi\left(x,t\right)
 \partial^{\mu}\phi\left(x,t\right).
\end{eqnarray}
Here $\mu=0,1$ and the parameters $M$ and $m$ denote the masses of the
fermionic and bosonic fields, respectively. We choose the following
representation for the Dirac matrices: $\gamma^0=\sigma_1$,
$\gamma^1=i\sigma_3$, $\gamma^5=\gamma^0\gamma^1=\sigma_2$. Then the
equations of motion become
\begin{equation}\label{e1}
i\sigma_1\partial_t\psi - \sigma_3\partial_x\psi - M \left[\cos
\phi\left(x,t\right)+i\sigma_2 \sin \phi\left(x,t\right)\right]\psi
=0,
\end{equation}
and
\begin{equation} \vspace{.2cm}\label{e2}
\partial_t\partial^t\phi\left(x,t\right) + \partial_x\partial^x
\phi\left(x,t\right)+i M{\bar{\psi}} \sigma_2
\textit{e}^{i\phi\left(x,t\right)\sigma_2}\psi + \lambda \phi^3
\left(x,t\right)- m^2\phi\left(x,t\right) =0,
\end{equation}
where \begin{equation}\label{e31} \psi=\left(\! \begin{array}{c}
\psi_{1}\\
\psi_2
\end{array}\! \right).
\end{equation}
Let us define
\begin{equation}\label{e3}
\xi=e^{-iEt} \left(\! \begin{array}{c}\xi_1+i\xi_2\\
\xi_3+i\xi_4\end{array}\! \right)=\left(\! \begin{array}{c}\psi_1+i\psi_2\\
\psi_1-i\psi_2\end{array}\! \right).
\end{equation}
We are interested in the solitary wave solution of the $\phi^4$ theory, in
which $\partial_t\phi=0$. Inserting Eq.\,(\ref{e3}) into
Eqs.\,(\ref{e1}) and (\ref{e2}) and using $\partial_t\phi=0$, we
obtain
\begin{align}\label{e4}
-& \partial_x\partial_x \phi\left(x\right)-M\cos \phi\left(x\right)\left(\xi_1 \xi_3+\xi_2\xi_4\right)\nonumber\\
+&M\sin \phi\left(x\right)\left(\xi_1\xi_4-\xi_2\xi_3\right)+\lambda \phi^3\left(x\right) -m^2 \phi\left(x\right)=0,\\
&\xi^{'}_1+M\cos \phi\left(x\right)\xi_3-E\xi_2-M\sin \phi\left(x\right)\xi_4=0,\\
&\xi^{'}_2+M\cos \phi\left(x\right)\xi_4+E\xi_1+M\sin
\phi\left(x\right)\xi_3=0,\\
&\xi^{'}_3+M\cos \phi\left(x\right)\xi_1+E\xi_4+M\sin
\phi\left(x\right)\xi_2=0,\\\label{e42} &\xi^{'}_4+M\cos
\phi\left(x\right)\xi_2-E\xi_3-M\sin \phi\left(x\right)\xi_1=0,
\end{align}
where prime denotes differentiation with respect to $x$.

 This set of equations cannot be solved analytically.
Therefore, we use a numerical method called the relaxation method
which determines the solution by starting with a guess and improving
it iteratively to relax to the true solution \cite{newman}. In order
to have a rapid convergence, we need a good initial guess which is
close to the true solutions and it also should satisfy the same
boundary solitons as the real solutions. Thus, we choose the fermionic
bound states of the system solved by Gousheh and Mobilia \cite{dr}
as the initial trial solution. Rather than a coupled system in which
both $\psi$ and $\phi$ are treated as dynamical fields, they studied
the spectrum of the fermion in the presence of a prescribed
$\phi\left(x\right)$. They considered an interacting system of a
fermion and a pseudoscalar field $\phi\left(x\right)$ described by
the Lagrangian
\begin{equation}\label{e5}\vspace{.2cm}
  {\cal L} = {\bar{\psi}}\left(i \gamma^\mu \partial_\mu
 - M e^{i\phi\left(x\right)\gamma^5}\right)\psi .
\end{equation}
They chose $\phi(x)$ to be piecewise linear as shown in
Fig.\,(\ref{fig.1}) by the dashed line. This is qualitatively similar to the kink
of the $\phi^4$ theory (dotted line in (\ref{fig.1})) or the soliton
of the sine-Gordon theory. The eigenfunctions of the Hamiltonian,
i.e. the solutions to the equations (\ref{e4}-\ref{e42}), can be chosen
to be also the eigenfunctions of the parity operator since the
Hamiltonian is invariant under parity. In the previous
representation the parity operation on the fermionic field is given
by $P\psi\left(x,t\right)=\sigma_1 \psi\left(-x,t\right)$. In the
representation (\ref{e3}), it becomes
$P\xi\left(x,t\right)=-\sigma_2 \xi\left(-x,t\right)$. Since we
shall choose our solutions to be eigenstates of the parity operator,
it is sufficient to solve the equations only for $x\geq0$. Also, we
map the $x$-interval $[0,\infty)$ to $[0,1]$ by $X=\tanh(x)$. In the
language of the relaxation method there are seven coupled
first-order ODEs: one second-order, four first-order equations and
one equation for the eigenvalue which is the fermion energy.
Therefore, we need seven conditions at two boundaries $X=0$ and
$X=1$. At $X=0$ we choose three conditions: one parity condition,
$\phi(0)=0$ and one giving a value to one of the $\xi_i(0)$s. This
value is allowed to change so as to normalized $\psi(x)$. This is
extremely important since the equation for $\phi(x)$ is non-linear
both in $\phi$ and $\psi$. At $X=1$ we choose four conditions: three
of the $\xi_i(1)$s are set to zero and $\phi(1)=\theta_0$ (in most
cases we set $\theta_0=\pi$, in order to have a soliton with winding
number one). In addition to these boundary conditions, we have to
add another condition to obtain a proper solitonic solution for
$\phi(x)$, i.e. one with finite energy and localized energy density. Since in
 our model $U(\phi)=\frac{{1}}{{4}}\lambda \left(\phi^2\left(x,t\right) -
\frac{m^2}{\lambda}\right)^2$, we have to set $\theta_0=\frac{m}{\sqrt{\lambda}}$
 in all cases in order to obtain a
solitonic solution for $\phi(x)$.
\section{Results}
To simplify the numerical calculations, we rescale the quantities of
the system by dividing Eq.\,(\ref{e1}) by $M^{3/2}$ and
Eq.\,(\ref{e2}) by $M^2$. Thus, the scaled equations of motion are
in the following form
\begin{equation}\label{r1}
i\sigma_1\partial_{t'}\psi' - \sigma_3\partial_{x'}\psi' -
\left[\cos \phi'\left(x',t'\right)+i\sigma_2 \sin
\phi'\left(x',t'\right)\right]\psi' =0,
\end{equation}
and
\begin{align} \vspace{.2cm}\label{r2}\vspace{.2cm}
\partial_{t'}\partial^{t'}\phi'\left(x',t'\right) &+
\partial_{x'}\partial^{x'}
\phi'\left(x',t'\right)+i {\bar{\psi'}} \sigma_2
\textit{e}^{i\phi'\left(x',t'\right)\sigma_2}\psi'
\hspace{.3cm}\nonumber\\\hspace{.3cm}&+ \lambda' \phi'^{3}
\left(x',t'\right)- m'^2\phi'\left(x',t'\right) =0.
\end{align}
Now all of the quantities are dimensionless and are given by
$t'=Mt$, $x'=Mx$, $\psi'=\frac{\psi}{\sqrt{M}}$, $\phi'=\phi$,
$m'=\frac{m}{M}$ and $\lambda'=\frac{\lambda}{M^2}$.\vspace{.2cm}
From now on we drop the primes for simplicity. In
most of the results presented here, we compare three models. The
first model which we call model I is our model, i.e. an eigenvalue
problem for a fermion chirally coupled to a pseudoscalar field which
is not prescribed and its shape, along with the functional forms of
the fermionic states, are determined by the equations of motion. The
second model which we refer to as model II is similar to our model,
except that in this model the pseudoscalar field coupled to fermion
is precisely the kink as a prescribed background field and cannot be
altered by the presence of the fermion. The third one which we refer
to as model III is the exactly solvable model solved by Gousheh and
Mobilia, in which again the pseudoscalar field is prescribed but in
the form of a piecewise linear field.
\par
\begin{figure}[th] \hspace{4.5cm} \includegraphics[width=8cm]{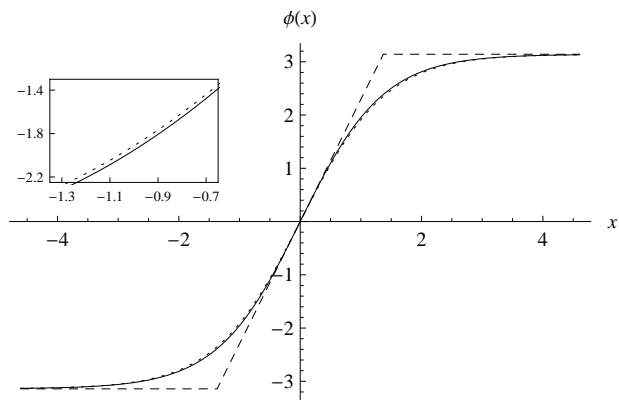}\caption{\label{fig.1} \small
  The graphs of the $\phi\left(x\right)$ for three cases: solid line depicts the $\phi\left(x\right)$ obtained
   from our model (model I), dotted line the prescribed kink solution of the pure $\phi^4$ theory (model II), and dashed
   line the prescribed background field of the exactly solvable model (model III).}
\end{figure}

 In Fig.\,(\ref{fig.1}) we show $\phi\left(x\right)$ for $m=1$ and
$\theta_0=\pi$ for three cases: the one obtained dynamically for our
system (solid line), the prescribed kink solution (dotted line) and
the prescribed form chosen in \cite{dr} (dashed line). It can be
seen that the solitonic solution is affected very slightly by the
fermion; the soliton in the presence of the fermion is so similar to
the isolated kink that they cannot be easily distinguished from each
other. Therefore, the back reaction of the fermion on the soliton is
indeed very small.

\par
\begin{center}
\begin{figure}[th] \hspace{3cm}\includegraphics[width=8cm]{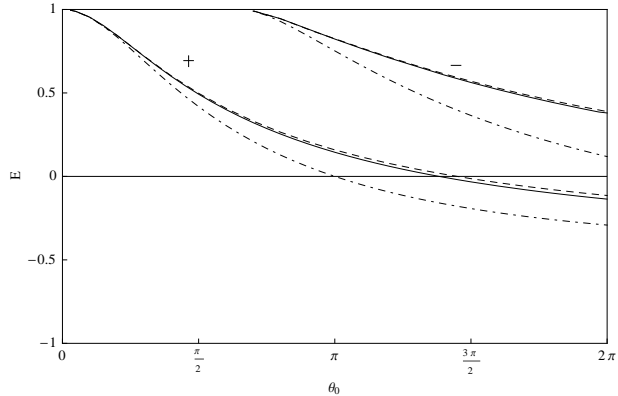}\caption{\label{fig.3} \small
   The fermionic bound state energies when the slope of the pseudoscalar fields at
    $x=0$ is $\mu\approx 2.957$, in the interval $0\leqslant\theta_0\leqslant2\pi$. Solid, dashed and dotdashed
    lines depict the fermion energies for the models I, II and III, respectively. The plus and minus signs indicate
    the parity of the bound states.}
\end{figure}
\end{center}

\begin{figure}[th] \hspace{1cm}\includegraphics[width=13cm]{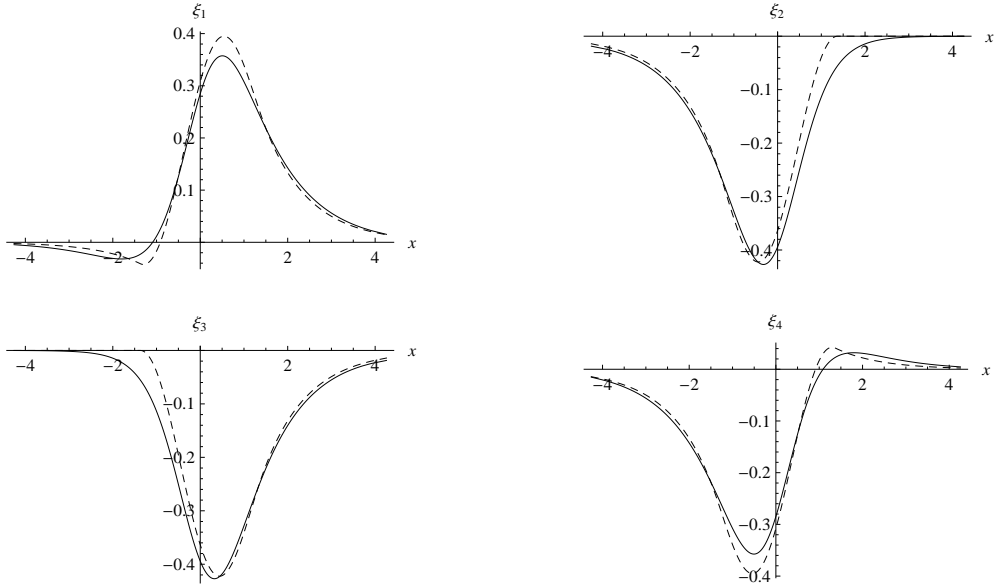}\caption{\label{fig.2} \small
  The lowest bound state of the fermion as a function of $x$. Solid curves are
  the solutions to model I and dashed lines the solutions to the model III.
   The parity is positive and both the fermion and the boson have unit masses.}
\end{figure}

 In Fig.\,(\ref{fig.3}) we show the fermionic bound state energies in the interval $0\leqslant\theta_0\leqslant2\pi$ for the model I (solid
lines), the model II (dashed lines) and the model III (dotdashed
lines). In this figure the slope of the pseudoscalar field at $x=0$
is fixed at about $2.957$ for all three models. This slope is
important in model III, where the critical point of a bound state
crossing $E=0$ occurs at $\theta_0=\pi$. In the models I and II this
crossing would happen for a greater value of $\theta_0$. Therefore,
the domain of the adiabatic region is larger in the relaxed state and the
isolated kink which is similar to it.
This result can also be illustrated as follows. If we insist that
the energy level crossing occurs at $\theta_0=\pi$ in all cases, the
slope of the soliton with winding number one would be greater in
models I and II. Therefore, the domain of the adiabatic region as
measured by the slope is larger. Comparison between the energies of
the models I and II shows that choosing kink as a prescribed
background is a very good approximation for our system.
From now on we set $\theta_0=\pi$ in all cases. Figure (\ref{fig.2})
shows $\xi_i$s for the lowest bound state for the case $m=1$. Solid
lines refer to the results obtained numerically for our lagrangian
(model I) and dashed lines refer to the solutions of the
equations of the Lagrangian (\ref{e5}) in which $\phi(x)$ is the
piecewise linear form shown by the dashed line in
Fig.\,(\ref{fig.1}) (model III). Since there are very small and
indistinguishable differences between the solutions of the models I
and II, we do not present the results of the model II.

\begin{figure}[th] \hspace{3.cm}\includegraphics[width=8cm]{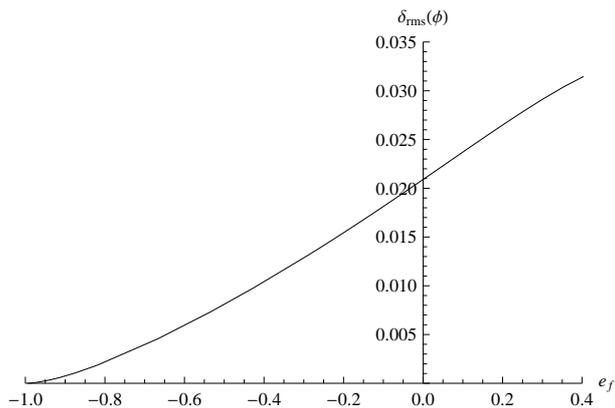}\caption{\label{fig.21} \small
  This is a plot of the point-wise root mean squared difference between the relaxed and the isolated kink. The independent variable is the energy of
  the fermionic bound state which is coupled to the soliton and is chosen to be the lowest bound state. This axis could equally have been chosen to be the mass of the elementary boson ($m$). The correspondence between the two is that $e_f=-1(M)\leftrightarrow m=\infty$ and $e_f=0.4(M)\leftrightarrow m\approx0.8(M)$. The numerical results for $\delta_{\textit{rms}}(\phi)$ becomes unreliable, when mass of the elementary boson is smaller than $0.6(M)$.}
\end{figure}

\begin{figure}[th]\hspace{4cm} \includegraphics[width=8cm]{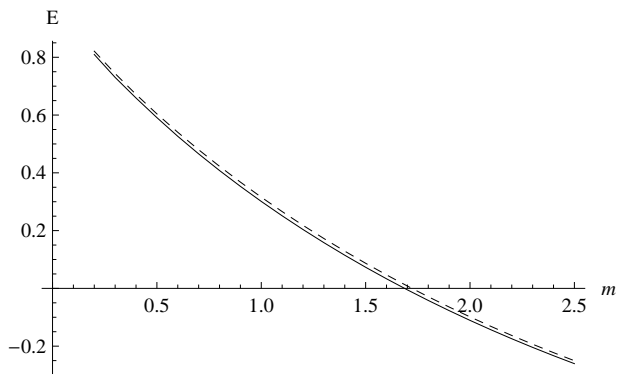}\caption{\label{fig.4} \small
  The energy of the lowest bound state of the fermion as a function of the boson mass for
   the model I (solid line) and model II (dashed line) at $\theta_0=\pi$.}
\end{figure}

\begin{figure}[th]\hspace{4cm} \includegraphics[width=8cm]{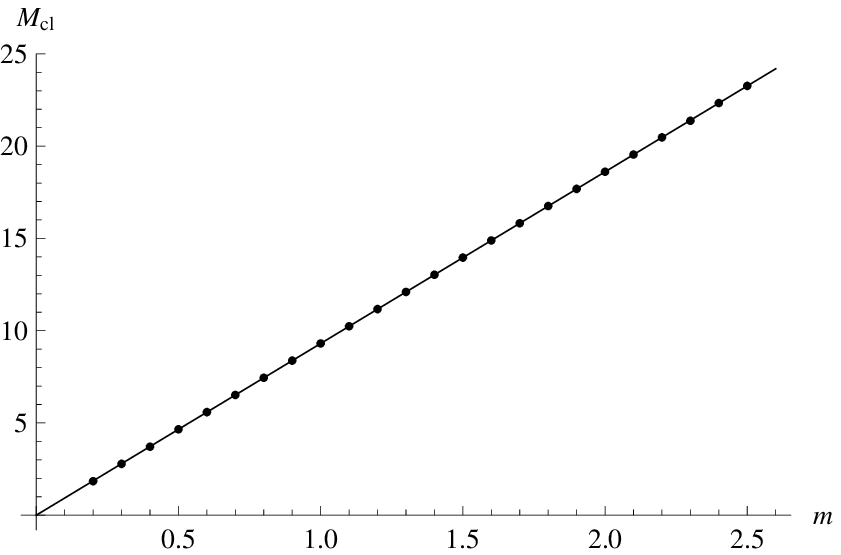}\caption{\label{fig.5} \small
  The total energy of the soliton as a function of the boson mass. The points show
   the numerical results for the model I and solid line shows the kink mass which
    behaves as $M_{cl}=\frac{2\sqrt{2}}{3}\theta_0^2 m$.}
\end{figure}

Figure (\ref{fig.4}) shows the energy of the lowest bound state of
the fermion as a function of the boson mass.
 In this graph we also show the fermion energy
of the model II. Note that, as expected, the two energies are close
and the relaxed one (model I) is slightly lower.
 The total boson energy, sometimes called the classical
soliton mass $M_{cl}$, as a function of the boson mass is presented
in Fig.\,(\ref{fig.5}). Note that for an isolated kink, the
functional form would be exactly linear as given by
$M_{cl}=\frac{2\sqrt{2}}{3}\theta_0^2 m$, as depicted in the figure
by the solid line. In this figure the result of the model I is shown
by the points to make these two results (models I and II)
distinguishable. It is interesting to note that the two results
are too close to each other to be distinguished. We can quantify the difference between the functional forms
of the relaxed and the isolated kink by computing their point-wise root mean squared difference. We have plotted this quantity as a function of the energy of the lowest fermionic bound state which is coupled to the soliton in Fig.\,(\ref{fig.21}). As it is evident from the graph, when mass of the elementary boson and therefore the soliton is infinite, the shape of the soliton is completely unaffected by the presence of the fermion and $\delta_{\textit{rms}}(\phi)\rightarrow 0$. As the mass of the boson is decreased, its sensitivity to the presence of the fermion increases. An important point is worth mentioning. It has been discovered as early as 1976 \cite{jakiew,backreaction1,backreaction2} that when the problem possesses particle conjugation symmetry, the zero fermionic mode has absolutely no back reaction on the soliton. In fact one can show on general grounds (using the definition of this operator, its unitarity property and general properties of the $\gamma$-matrices) that when the problem possesses this symmetry, the zero modes produce no back reaction. However, our problem does not have this symmetry and the effect of the zero mode on the shape of the soliton is non-zero. As a matter of fact Fig.\,(\ref{fig.21}) shows that the pattern of $\delta_{\textit{rms}}(\phi)$ does not show any irregularity at $e_f=0$.
\par In the Skyrme model
\cite{skyrme3} the pions are introduced as the pseudoscalar
background field, coupled to baryons playing the role of fermions.
Skyrme conjectured that the solitonic excitations of pions, later on
called skyrmions, are indeed nucleons. This conjecture has been
studied by many authors and used as an approximate phenomenological
model at energy scales up to about a few $Gev$ (see for example
\cite{conjec1,conjec2,conjec3,conjec4,conjec5}).
 Our results are also consistent with this conjecture. By extrapolating
 the results displayed in Fig.\,(\ref{fig.5}) for
 our model, one can conclude that when $M_{cl}=1$, the elementary
 boson mass is $m\approx0.116$, all in units of the fermion mass $M$.
  For the soliton with $M_{cl}=1$, $\mu\approx 0.258$ which corresponds
to a wide soliton with winding number one. Also, as can be seen from
Fig.\,(\ref{fig.4}), at $m\approx0.116$ there is no fermion energy
level crossing zero. Therefore, here the soliton of mass $M_{cl}=1$
polarizes the vacuum and its ground state has fermion number one.
The pion mass is about $0.14 Gev$ and the nucleon mass is about $1.0
Gev$. Therefore, if in our model the pseudoscalar field $\phi(x)$
plays the role of the pions, solitonic excitations (with $M_{cl}=1$
for $m\approx0.116$) could be considered to be nucleons. Note that
in the Skyrme model there are three pions introduced as in a
non-linear sigma model and two doublets for two nucleons, but in our
model there are only one real scalar field and one fermion field.
This could partly account for the difference in the results.

\par
\begin{figure}[th]\hspace{3.3cm} \includegraphics[width=8cm]{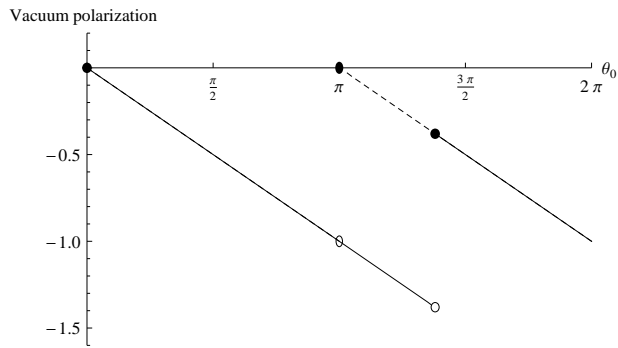}\caption{\label{fig.8} \small
The vacuum polarization by the soliton as a function of $\theta_0$ when the slope of the pseudoscalar field at
 $x=0$ is $\mu\approx 2.957$. The solid lines show the result for our model and the dashed lines the result for the model III. The filled and empty circles
 at $\theta_0\approx 1.38 \pi$ are for model I and the filled and empty ellipses at $\theta_0=\pi$ are for model III.}
\end{figure}

 Figure (\ref{fig.8}) shows the vacuum polarization induced by the soliton as a function of $\theta_0$
when the slope of the pseudoscalar field at
    $x=0$ is $\mu\approx 2.957$. Solid lines show the result of our model (model I) in which the fermion energy level crosses $E=0$ at $\theta_0\approx 1.38 \pi$ and the dashed lines show the result of the model III
    in which the fermion energy level crosses $E=0$ at $\theta_0=\pi$. There are two contributions to the vacuum polarization.
First is the adiabatic contribution, predicted by Goldstone and Wilzcek \cite{goldstone}, which is a linear decrease with slope $-\theta_0/\pi$
whenever the interaction is in the form of $\bar{\psi} \textit{e}^{i\phi\gamma^5}\psi$.
This is due to the occurrence of the spectral deficiency in the Dirac sea as $\theta_0$ increases. 
The second is the non-adiabatic contribution which occurs only when a fermionic energy level crosses $E=0$. 
In the latter case there is a jump of $+1$ in the vacuum polarization since this energy level is filled after crossing $E=0$ in the vacuum state, by definition. These two contributions have been explicitly
derived for model III in reference \cite{dr}. There is a potential ambiguity of whether the state at $E=0$ is to be considered filled or empty in the vacuum state. 
If this is considered as a true ambiguity, one can resolve it by observing that the vacuum polarization has to be zero at $\theta_0=0$,
 and self-consistency and symmetry dictates the situation at the other points as shown in Fig.\,(\ref{fig.8}). 
We should mention that in our model which is partially based on $\lambda \phi^4$ theory, we have only two distinct topological sectors: the trivial and the topological charge $\pm 1$ sectors.
However, the fermion responds to the value of $\theta_0$, regardless of the topological charge.
\par
\begin{figure}[th] \hspace{4cm}\includegraphics[width=8cm]{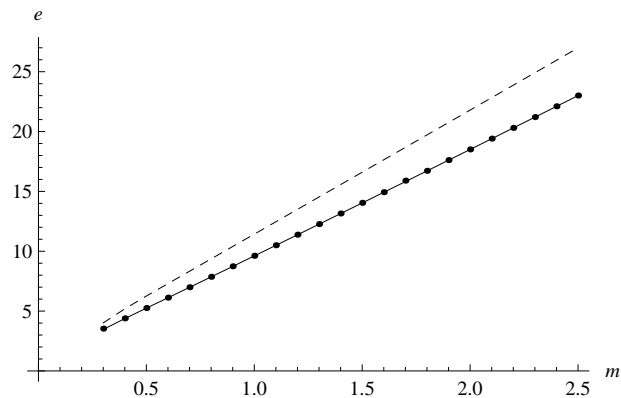}\caption{\label{fig.6} \small
  The total energy (the lowest fermionic bound state energy plus
the soliton energy) as a function of the boson mass. Dashed line
represents the total energy for the model III, and solid line represents the total energy for the model II and the points the total energy for our model.}
\end{figure}

 
In Fig.\,(\ref{fig.6}), we show the total energy, i.e. the lowest
fermionic bound state energy plus the soliton energy for both models I and III. 
After obtaining the solution of our model for each value of $m$, we set the slope $\mu$ of model III to the slope of $\phi(x)$ obtained at $x=0$. 
In this way we can compare the energies of two models for each $m$. 
As expected, the total energy for the relaxed kink (model I) is lower than for
the model III.

\begin{figure}[th]\hspace{4cm} \includegraphics[width=8cm]{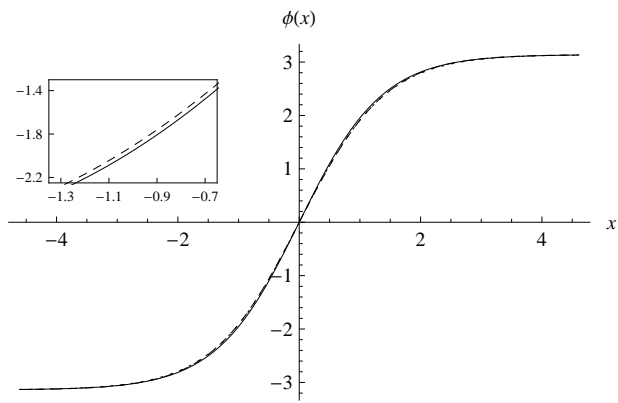}\caption{\label{fig.7} \small
   $\phi (x)$ for two cases: solid line shows the soliton relaxed along with the lowest
    fermionic bound state and dashed line shows
   the soliton relaxed along with the first excited state.}
\end{figure}

In Fig.\,(\ref{fig.7}) we compare $\phi (x)$ for two cases: The
soliton relaxed with the ground state and the first excited state in
which the parities are positive and negative, respectively. 
The parameters of the system are the same for both cases: $m=1$ and
$\theta_0=\pi$. As can be seen from the graph, the relaxed solitons of these
two states are not the same.

\section{Conclusion}
In this paper, we have considered a fermion field chirally
coupled to a pseudoscalar field which could evolve into a soliton in
(1+1) dimensions. We have solved the system self-consistently and
found the lowest lying bound states of the fermion and the shape of the
soliton non-perturbatively and exactly to within our numerical
approximation. We have also presented the results of a
similar model in which the pseudoscalar field is the prescribed kink
instead of a field which can be altered due to the presence of
the fermion. Comparing the results of this model, including those of the fermionic states, with our system, we find that we can
approximate the exact (relaxed) kink by the isolated kink, as is
usually done.
 We have shown that in the models with
the relaxed kink and the prescribed kink, the adiabatic region is larger than
the model with the prescribed piecewise linear background field as a soliton. We have also computed
 the vacuum polarization of the fermion induced by the presence of
  the soliton and displayed the results for our model and the exactly solvable prescribed model.
   In this connection we have elaborated on the adiabatic and non-adiabatic contributions to the vacuum polarization. In addition, we have
determined that when the soliton mass $M_{cl}$ is equal to the fermion
mass $M$, the elementary boson mass $m\approx0.116 M$ which is in a good
agreement with the conjecture of Skyrme. Moreover, we have shown
that the relaxed form of the soliton depends on the fermionic state
to which it is coupled. It is worth mentioning that we have not
considered the back reaction perturbatively but exactly to within
the numerical approximation.

 \end{document}